\documentclass{emulateapj}
\usepackage{amsmath}

\shorttitle{The mass-loss rates of WC and WO stars}
\shortauthors{Tramper, Sana, \& de Koter}

\begin{document}

\title{A new prescription for the mass-loss rates of WC and WO stars}

\author{F. Tramper\altaffilmark{1}}
\affil{European Space Astronomy Centre (ESA/ESAC), Operations Department, Villanueva de la Ca\~nada (Madrid), Spain}
\email{ftramper@sciops.esa.int}

\author{H. Sana}
\affil{Institute of Astrophysics, KU Leuven, Celestijnenlaan 200 D, B-3001, Leuven, Belgium}

\and

\author{A. de Koter}
\affil{Anton Pannekoek Institute for Astronomy, University of Amsterdam, PO Box 94249, 1090 GE Amsterdam, The Netherlands}
\affil{Institute of Astrophysics, KU Leuven, Celestijnenlaan 200 D, B-3001, Leuven, Belgium}

\altaffiltext{1}{ESA Research Fellow}

\begin{abstract}

We present a new empirical prescription for the mass-loss rates of carbon and oxygen sequence Wolf-Rayet stars as a function of their luminosity, surface chemical composition, and initial metallicity.  The new prescription is based on results of detailed spectral analyses of WC and WO stars, and improves the often applied \cite{nugis2000} relation. We find that the mass-loss rates of WC and WO stars (with $X=0$ and $Y \la 0.98$) can be expressed as 
\begin{center}
$\log{\dot{M}} = -9.20 + 0.85\log{(L/L_{\odot})} + 0.44\log{Y} + 0.25\log{(Z_{\mathrm{Fe}}/Z_{\mathrm{Fe}, \odot})}$.\\
\end{center}

\noindent This relation is based on mass-loss determinations that assume a volume-filling factor of 0.1, but the prescription can easily be scaled to account for other volume-filling factors. The residual of the fit is $\sigma = 0.06$ dex. We investigated whether the relation can also describe the mass loss of hydrogen-free WN stars and showed that it can when an adjustement of the metallicty dependence ($\log{\dot{M}} \propto 1.3\log{(Z_{\mathrm{Fe}}/Z_{\mathrm{Fe}, \odot})}$) is applied. Compared to \cite{nugis2000}, $\dot{M}$ is less sensitive to the luminosity and the surface abundance, implying a stronger mass loss of massive stars in their late stages of evolution. The modest metallicity dependence implies that if WC or WO stars are formed in metal deficient environments, their mass-loss rates are higher than currently anticipated. These effects may result in a larger number of type Ic supernovae and less black holes to be formed, and may favour the production of superluminous type Ic supernovae through interaction with C and O rich circumstellar material or the dense stellar wind. 

\end{abstract}


\section{Introduction}

\begin{table*}
\begin{center}
\caption{Spectral types and parameters of the calibration stars.\label{tab:calstars}}
\begin{tabular}{l l l l c c c c l}
\tableline\tableline
\#\tablenotemark{a} & ID		& Alt. ID & SpT & $\log{(L/L_{\odot})}$	& $\log{\dot{M}}$	& $Y$	& $Z_{\mathrm{Fe}}/Z_{\mathrm{Fe}, \odot}$		& Reference \\
\tableline
1 & WR11\tablenotemark{b}	& $\gamma$ Vel		& WC8	& 5.0		& $-$5.1	& 0.64	& 1.0		& \cite{demarco2000} \\
2 & WR26	&					& WN7/WCE	& 6.1	& $-$4.01	& 0.80	& 1.0		& \cite{sander2012} \\
3 & WR58	&					& WN4/WCE	& 5.15& $-$4.80& 0.975	& 1.0		& \cite{sander2012} \\
4 & WR90	& HD 156385			& WC7	& 5.5		& $-$4.6	& 0.53	& 1.0		& \cite{dessart2000} \\
5 & WR93b	&					& WO3	& 5.30	& $-$5.00	& 0.29	& 1.0		& \cite{tramper2015} \\
6 & WR102	&					& WO2	& 5.45	& $-$4.92	& 0.14	& 1.0		& \cite{tramper2015} \\
7 & WR103	& HD 164270			& WC9	& 4.9		& $-$5.0	& 0.61	& 1.0		& \cite{crowther2006a} \\
8 & WR111	& HD 165763			& WC5	& 5.3		& $-$4.8	& 0.38	& 1.0		& \cite{hillier1999} \\
9 & WR135	& HD 192103			& WC8	& 5.2		& $-$4.9	& 0.66	& 1.0		& \cite{dessart2000} \\
10 & WR142	& 					& WO2	& 5.39	& $-$4.94	& 0.26	& 1.0		& \cite{tramper2015} \\
11& WR145	&					& WN7/WCE	& 5.8	 & $-$4.35	& 0.935	& 1.0		& \cite{sander2012} \\
12& WR146\tablenotemark{b}	& 					& WC5	& 5.7		& $-$4.5	& 0.76	& 1.0		& \cite{dessart2000} \\
13 & Brey 7		& HD 32125, BAT 9		& WC4	& 5.44	& $-$4.8	& 0.65	& 0.5		& \cite{crowther2002} \\
14 & Brey 8		& HD 32257, BAT 8		& WC4	& 5.42	& $-$4.9	& 0.45	& 0.5 	& \cite{crowther2002} \\
15 & Brey 10	& HD 32402, BAT 11		& WC4	& 5.70	& $-$4.5	& 0.66	& 0.5		& \cite{crowther2002} \\
16 & Brey 43	& HD 37026, BAT 52		& WC4	& 5.65	& $-$4.5	& 0.46	& 0.5		& \cite{crowther2002} \\
17 & Brey 50	& HD 37680, BAT 61		& WC4	& 5.68	& $-$4.4	& 0.74	& 0.5		& \cite{crowther2002} \\
18 & Brey 74	& HD 269888, BAT 90	& WC4	& 5.44	& $-$4.8	& 0.45	& 0.5		& \cite{crowther2002} \\
19 & Brey 93	& BAT 123			& WO3	& 5.20	& $-$5.14	& 0.30	& 0.5		& \cite{tramper2015} \\
20 & $[$L72$]$LH41-1042 &					& WO4	& 5.26	& $-$5.05	& 0.22	& 0.5		& \cite{tramper2015} \\
21 & DR 1 in IC 1613		&					& WO3	& 5.68	& $-$4.76	& 0.44	& 0.15	& \cite{tramper2013} \\

\tableline
\end{tabular}
\tablenotetext{1}{Used as labels in Figures~\ref{fig:WR_mdot} and \ref{fig:mdot_Y}}
\tablenotetext{2}{WR + O binary. The contribution from the O star was taken into account in the spectroscopic analysis.}
\end{center}
\end{table*}

Classical Wolf-Rayet (WR) stars are evolved massive stars characterised by their dense, optically thick outflows to which they own their tell-tale emission-line spectra. Driven by radiation pressure \citep[e.g., ][]{vink2005, grafener2005}, the winds of these stars require an efficient momentum transfer from the radiation to the gas, usually quantified by means of the wind performance number $\eta \equiv \dot{M}v_{\infty}/(L/c)$ (with $\dot{M}$ the mass-loss rate, $v_{\infty}$ the terminal velocity of the wind, and $L$ the stellar luminosity). For WC and WO stars, $\eta$ reaches values of around 10 \citep[e.g., ][]{sander2012, tramper2015}, and most WN stars have $\eta$ close to unity \citep[e.g., ][]{hamann2006, hainich2014, hainich2015}. This indicates that multiple photon scatterings are required to drive the wind.

A good empirical knowledge of the outflow properties during the Wolf-Rayet phase is necessary to understand the mechanism that drives their winds, as well as to provide accurate values to be used in evolutionary models. The mass-loss efficiency in the WR phase strongly impacts the immediate pre-supernova evolution, and determines the predicted type of supernova as well as the type of compact object that is produced. 

\citet[][henceforth NL00]{nugis2000} provide an empirical mass-loss prescription for Wolf-Rayet stars as a function of luminosity and surface chemical composition. The NL00 rates are currently implemented in most evolutionary models, either using the separate WN and WC prescriptions \citep[Equations 20 and 21 in NL00, e.g., in the Geneva models;][]{ekstrom2012} or the combined WR presciption \citep[Equation 22 in NL00, e.g., in {\sc mesa};][]{paxton2011}. However, the mass-loss rates of the oxygen-sequence Wolf-Rayet stars, which have a very low surface helium abundance, cannot be reproduced by the NL00 prescriptions \citep{tramper2015}.

In this paper we provide a new prescription for the mass-loss rates of hydrogen-free Wolf-Rayet stars, significantly improving the predictions for the mass-loss rates during the WC and WO phases. The next Section presents the calibration of this new prescription. Its dependence on stellar parameters and the parameter domain in which it is valid is then discussed in Section~\ref{sec:discussion}. We summarise our findings in Section~\ref{sec:summary}.

\section{Calibration of WC and WO star mass loss}

The aim of this work is to revisit the prescription of WR mass-loss rates and to obtain a new calibration that is valid over the full parameter space covered by the WC and WO stars. To do this, we assume that the mass-loss rates ($\dot{M}$, in $M_{\odot}$ yr$^{-1}$) of these stars can be described by a relation of the form
\begin{align}
\log{\dot{M}} = A + B\log{\frac{L}{L_{\odot}}} + C\log{Y} + D\log{\frac{Z_{\mathrm{Fe}}}{Z_{\mathrm{Fe}, \odot}}}\label{eq:mdot}.
\end{align}

This relation is similar to the one of NL00, i.e. the mass loss depends on the stellar luminosity $(L)$ and the surface helium mass fraction $(Y)$. In the case of the hydrogen-free WR stars, the surface mass-fraction of heavier elements is by definition $Z = 1-Y$, and we do not include a separate term for this component\footnote{Inclusion of a $\log{Z}$ term does not yield a significantly better fit to the data, as verified by an F-test $(p(F)>0.8$). The $\log{Z}$ coefficient is not significantly different from 0. Such a fit further results in larger formal uncertainties in the derived parameters as a result of the strong correlation between $Y$ and $Z$.}. In WC and WO stars, $Z$ is effectively the sum of the high carbon and oxygen abundances, and the contribution from iron-like elements to this mass fraction is small. However, iron-group elements are expected to be the dominant wind drivers even at high carbon and oxygen abundances \citep[e.g., ][]{crowther2002, vink2005}. We therefore explicitly include this dependence on the iron mass fraction $Z_{\mathrm{Fe}}$ in Equation~\ref{eq:mdot}. This ensures that the derived prescription is valid for stars in different metallicity environments. We adopt the solar abundances from \citet[][i.e. $Z_{\odot} = 0.014$]{asplund2009}.

We use the parameters of the WC and WO stars listed in Table~\ref{tab:calstars} to derive the coefficients of Equation~\ref{eq:mdot}. These stars have been selected using the following criteria: 1) the spectrum was quantitatively analysed using non-LTE atmosphere models that include line-blanketing and account for wind clumping, and 2) the surface abundances of carbon and oxygen have been modelled in the analysis (i.e. no grid-based analyses where the abundances were fixed). All WC and WO stars in the sample were analysed using {\sc cmfgen} \citep{hillier1998}, and the WN/WC stars with the Potsdam Wolf-Rayet ({\sc powr}) code \citep{grafener2002, hamann2004}. The WN/WC stars are included to provide calibration points at high $Y$ values and to increase the sample size. Exclusion of these stars results in an essentially identical fit, but with larger error bars on the coefficients. Two of the Galactic WC stars (WR11 and WR146) are WR+O binaries. While this has been taken into account in the spectroscopic analysis, we note that their properties may be more uncertain.

All mass-loss rates in Table~\ref{tab:calstars} correspond to a volume-filling factor of $f_\mathrm{c} = 0.1$, which is typical for Wolf-Rayet stars. The predicted mass-loss rates are thus only valid if the winds of the calibration stars indeed have this volume filling factor, but can easily be scaled to account for other values of $f_{\mathrm{c}}$ (using $\dot{M} \propto f_{\mathrm{c}}^{0.5}$).

\begin{figure*}
\epsscale{0.7}
\plotone{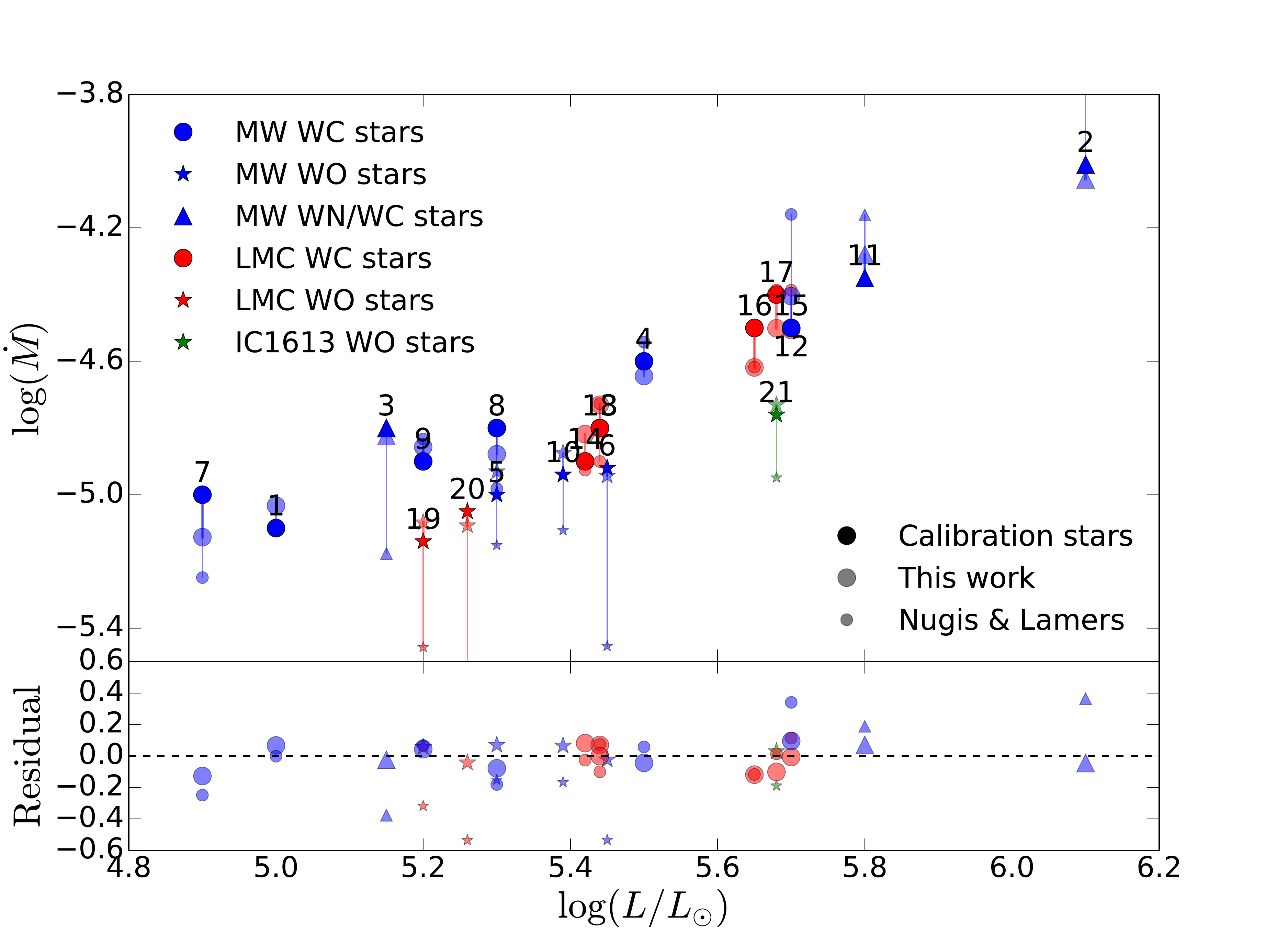}
\caption{Measured mass-loss rates versus luminosity of our sample stars (large solid symbols) compared to values computed by the prescription from this work (large transparent symbols) and NL00 (small transparent symbols). The bottom panel shows the residuals for both prescriptions.\label{fig:WR_mdot}}
\end{figure*}

Although not all spectral subtypes are represented, the sample of 21 stars provides a good coverage of the relevant parameter space, i.e. spanning a range of luminosities ($4.9 \leq \log{(L/L_{\odot})} \leq 6.1$; Figure~\ref{fig:WR_mdot}) and surface abundances ($0.14 \leq Y \leq 0.98$; Figure~\ref{fig:mdot_Y}). Twelve stars are located in the Milky Way (MW, $Z_{\mathrm{Fe}} = Z_{\mathrm{Fe}, \odot}$), and eight in the Large Magellanic Cloud (LMC, $Z_{\mathrm{Fe}} = 0.5 Z_{\mathrm{Fe}, \odot}$). The WO star DR1 located in IC1613 provides a valuable third metallicity point at $Z_{\mathrm{Fe}} = 0.15 Z_{\mathrm{Fe}, \odot}$.

The coefficients of Equation~\ref{eq:mdot} were derived by a linear regression to the data of Table~\ref{tab:calstars}. The resulting mass-loss prescription with 1$\sigma$ uncertainties on the coefficients is 
\begin{align}
\log{\dot{M}} = -9.20(\pm0.35) + 
0.85(\pm0.06)\log{\frac{L}{L_{\odot}}} \nonumber\\
+ 0.44(\pm0.08)\log{Y} + 
0.25(\pm0.08)\log{\frac{Z_{\mathrm{Fe}}}{Z_{\mathrm{Fe}, \odot}}}\label{eq:recipe}.
\end{align}

The mass-loss rates that are predicted by the new prescription are compared to the calibration values in Figures~\ref{fig:WR_mdot} and \ref{fig:mdot_Y}. The figures also show the mass-loss rates predicted by the NL00 prescription (their Equation~22) that is often applied in evolutionary models. The NL00 rates have been scaled with the theoretically predicted metallicity dependence of $\dot{M} \propto Z_{\mathrm{Fe}}^{0.66}$ \citep{vink2005}. 

The mass-loss rates from the new prescription match the calibration values better than the NL00 prescription. The most significant improvement is for the WO stars, where residuals decrease from 0.1-0.5 dex for NL00 to less than 0.05 dex for the new prescription. Over the whole calibration sample, the standard deviation of the mass-loss rates from the new recipe compared to the calibration values is $\sigma = 0.06$ dex, versus $\sigma = 0.18$ dex for NL00 rates ($\sigma = 0.17$ dex for the WC prescription of NL00, their Equation~21). The mean deviation of the mass-loss rates of the new recipe from the calibration values is $10^{-4}$ dex, indicating that there is no systematic offset to higher or lower mass-loss rates. These numbers indicate that the new prescription provides a significant improvement in the prediction of the mass-loss rates of WC and WO stars.


\section{Discussion}\label{sec:discussion}

In this Section we first compare our results to the sample of single Galactic WC stars from \cite{sander2012}. Then we discuss the implications of the change in dependencies of the mass-loss rates on luminosity, abundances, and metallicity for the late stages of evolution of massive stars.

\subsection{The Galactic WC stars}

\cite{sander2012} analysed all known presumed single Galactic WC stars using a grid of models from the Potsdam Wolf-Rayet ({\sc powr}) code. In the analysis, the surface mass fractions of helium, carbon, and oxygen were fixed to 0.55, 0.45, and 0.05, respectively. This may affect the derived luminosities and mass-loss rates for stars whose abundances deviate significantly from these values. In Figure~\ref{fig:mdot_Gal} we compare their results to the predictions from the new mass-loss prescription and those from the NL00 prescriptions. 

The four WC stars that are both in the calibration sample and in the \cite{sander2012} sample are marked in the plot. The considerable offset in luminosity and mass loss for WR90, WR103, and WR135 can be explained by differences in the adopted distances (0.8 kpc versus 1.55 kpc for WR90, 2.4 kpc versus 1.9 kpc for WR103, 1.7 kpc versus 1.4 kpc for WR135). However, the trend of mass-loss rate with luminosity is conserved for these stars (see also discussion below), all of which have helium abundances relatively close to $Y=0.55$ ($Y=0.53$ for WR90, $Y=0.61$ for WR103, and $Y=0.66$ for WR135). 

For the WC5 star WR111, the distances assumed in both analyses are comparable (1.55 kpc versus 1.6 kpc). Here, the offset between the derived luminosities and mass-loss rates may be a result of the assumed abundances in \cite{sander2012}, as the helium abundance derived by the detailed modelling is $Y=0.38$. If the other early-type WC stars have comparable abundances, this may explain the larger offset of these stars to the predictions from each of the prescriptions (see Figure~\ref{fig:mdot_Gal}, WC4 and WC5 subtypes marked with star symbols). 

\begin{figure*}
\epsscale{0.7}
\plotone{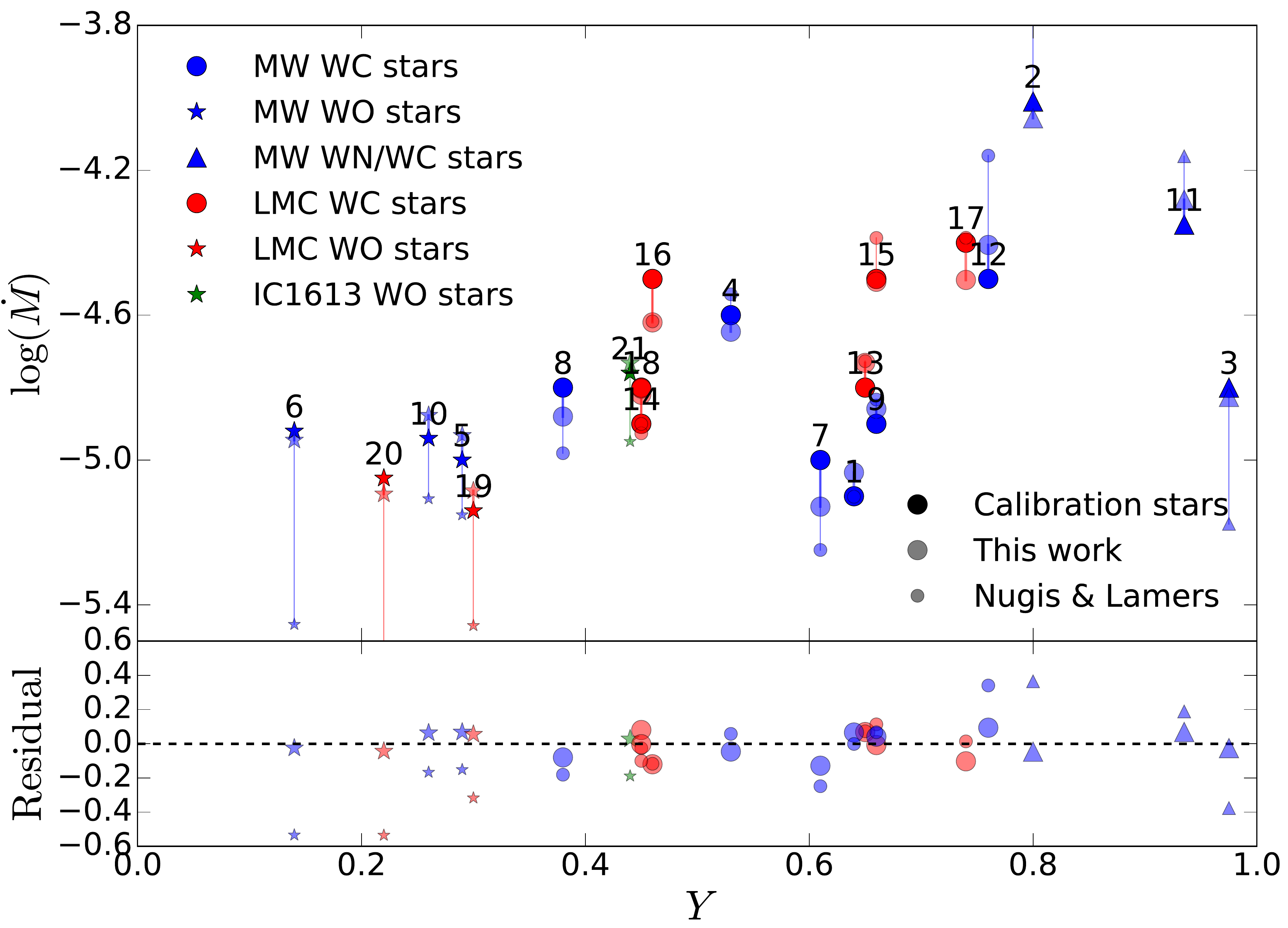}
\caption{Same as Figure~\ref{fig:WR_mdot}, but as a function of surface helium mass fraction.\label{fig:mdot_Y}}
\end{figure*}

\subsection{Dependencies and implications}

The dependence of the mass-loss rates of hydrogen-free WR stars on luminosity derived in this work ($\dot{M} \propto L^{0.85\pm0.06}$) is consistent with those derived for WC stars by NL00 ($\dot{M} \propto L^{0.84\pm0.17}$, their Equation~21) and \citet[][$\dot{M} \propto L^{0.83\pm0.11}$]{sander2012}.  This dependence can naturally be explained by the results of  \cite{sander2012}, who find that the transformed radius is proportional to the temperature squared. For fixed terminal wind velocities ($v_{\infty}$), this relation results in a $\dot{M} \propto L^{0.75}$ dependence. The slightly steeper dependence on $L$ is the result of an increasing $v_{\infty}$ towards earlier spectral subtypes \citep[see][for a discussion]{sander2012}. 

\begin{figure*}
\epsscale{0.7}
\plotone{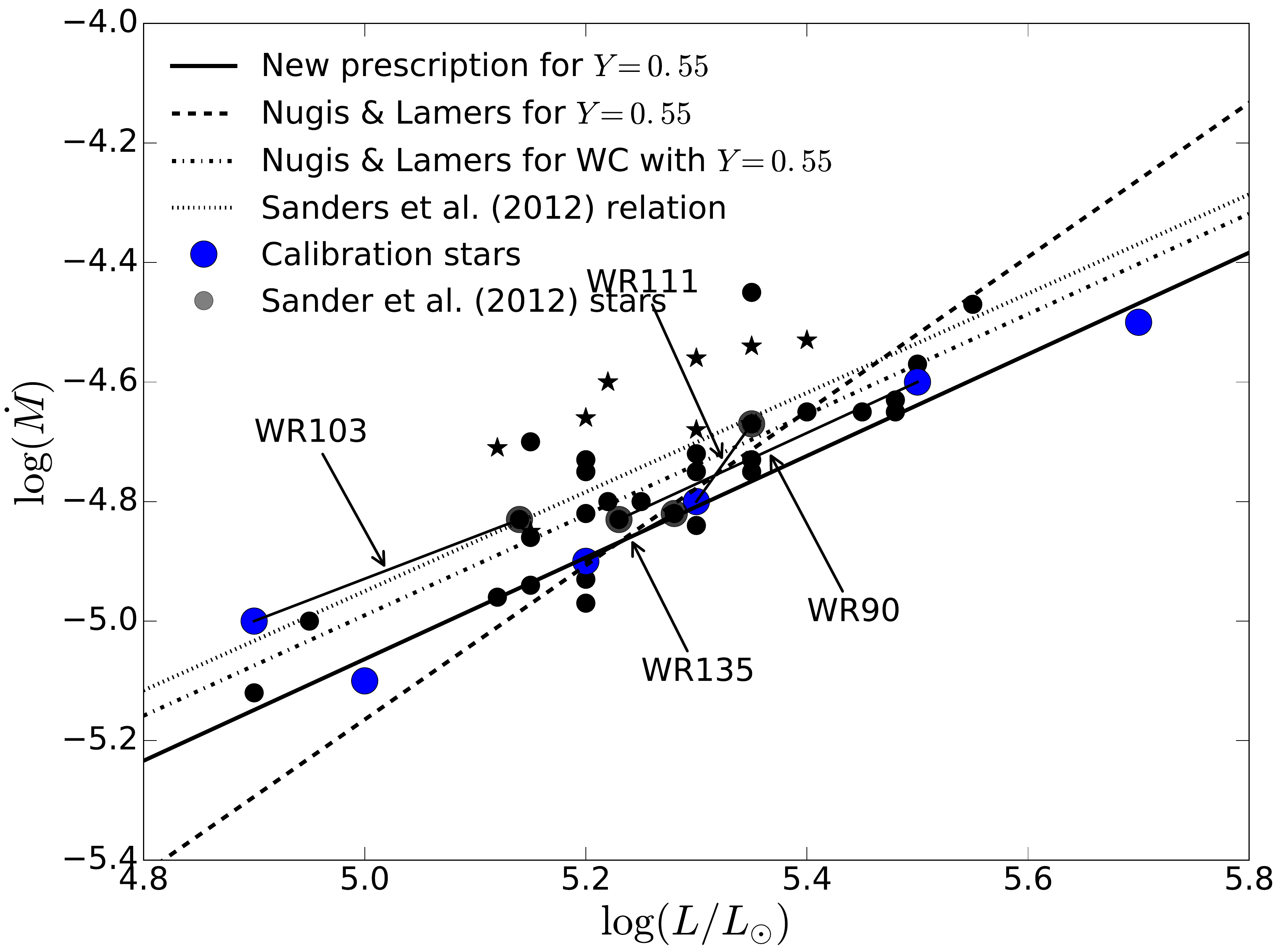}
\caption{Comparison of mass-loss rates from the new recipe and the NL00 recipes to the Galactic WC stars analysed by \cite{sander2012}. The mass-loss versus luminosity relation obtained by \cite{sander2012} is also indicated. The four WC stars that are in our calibration sample are indicated with the larger symbols, with the values from \cite{sander2012} in black and the values used in this work in blue.  The two calibration WC stars that are not in the \cite{sander2012} sample are also shown in blue symbols. Early-type WC stars (WC4 and WC5) are indicated with star symbols.\label{fig:mdot_Gal}}
\end{figure*}

The luminosity dependence of the combined WN+WC prescription of NL00 is much steeper ($\dot{M} \propto L^{1.29}$, their Equation~22). This results in an underprediction of the mass-loss rate for low luminosities, and an overprediction for high luminosities (see Figure~\ref{fig:mdot_Gal}). This effect reaches up to 0.2 dex in $\dot{M}$ for the lowest/highest luminosities. Implemented in stellar evolution models, these modifications in mass-loss properties may impact the surface abundance ratio of carbon and oxygen, which is essential in constraining the elusive $^{12}$C$(\alpha,\gamma)^{16}$O thermonuclear reaction rate \citep[e.g., ][]{grafener1998}. The NL00 prescription for WN stars also has a much steeper luminosity dependence ($\dot{M} \propto L^{1.63}$, their Equation~21). However, the transformed radius argument given above should hold for these stars, and this high value is likely a result of the very strong dependence on the helium abundance ($\dot{M} \propto Y^{2.22}$) in this prescription due to the inclusion of hydrogen-containing WN stars.

The largest impact on the late stages of evolution of massive stars comes from the dependence of the mass-loss rate on the surface abundances. The  combined NL00 prescription gives $\dot{M} \propto Y^{1.29}Z^{0.5}$. The abundance dependence of the NL00 WC relation is even stronger: $\dot{M} \propto Y^{2.04}Z^{1.04}$. The strong $Y$-dependence implies that the mass loss decreases strongly towards the late stages of evolution, an effect that is not in agreement with the derived mass-loss rates of the WO stars. The weaker $\dot{M} \propto Y^{0.44}$ dependence derived in this work does produce mass-loss rates in agreement with those derived for both the WC and WO stars. This implies that the mass-loss rates of WC and WO stars are considerably larger towards the later stages of evolution, where the surface helium abundance becomes low (see Figure~\ref{fig:mdot_Y}).

The combined effect of the luminosity and abundance dependencies is that more mass than currently predicted is lost during the last $\sim 40$\% of the core-helium burning phase and the post-helium burning phase. Implementation in evolutionary models is needed to assess the impact of this extra mass loss on the lifetimes of the WC and WO stars and on the properties of the direct progenitor stars of supernovae. Potential implications are that these properties favour an increased number of type Ic supernovae in either single star or close binary evolution. Our findings may also be relevant for the discussion on the nature of superluminous type Ic supernovae that appear to be associated with faint and metal-poor galaxies \citep{quimby2011, neill2011, chomiuk2011}. Among the scenarios proposed to explain these events \citep[see, e.g., ][for a discussion]{inserra2013} is the interaction of the supernova ejecta with a massive carbon- and oxygen-rich circumstellar medium \citep{blinnikov2010} or with the dense wind of the progenitor \citep{chevalier2011, ginzburg2012}. Furthermore, a higher mass loss in the hydrogen-free Wolf-Rayet phase would lead to a decrease in the number of black holes produced.

The metallicity dependence of the mass-loss prescription ($\dot{M} \propto Z_{\mathrm{Fe}}^{0.25}$) is also weaker than the one predicted by theory for WC stars \citep[$\dot{M} \propto Z_{\mathrm{Fe}}^{0.66}$,][]{vink2005}, as well as the one empirically derived from Galactic and LMC WC stars alone \citep[$\dot{M} \propto Z_{\mathrm{Fe}}^{\sim 0.5}$,][]{crowther2002}. To verify that the derived metallicity dependence is not dominated by the single calibration point at $0.15 Z_{\odot}$ we repeated the fit excluding this data point. This results in nearly identical values of coefficients $A$, $B$, and $C$ of Equation~\ref{eq:mdot} and their errors (small changes at the third decimal). The derived value of the metallicity dependence is $D = 0.20\pm0.12$, in good agreement with the value derived in Equation~\ref{eq:recipe}.

This weak dependence gives rise to stronger winds for WC and WO stars in low-metallicity environments than currently predicted, in agreement with the high mass-loss rate of the WO star in IC1613. However, the exposure of the deep layers with helium-burning products, necessary to produce WC and WO stars, depends on the mass-loss history in previous evolutionary stages, where the metallicity dependence of the stellar winds is found to be much higher (Equation~\ref{eq:recipeWN}). Thus, it is harder to form WC and WO type stars in low-metallicity environments, which may not be possible by mass loss through stellar winds alone. Instead, alternative mass-loss mechanisms such as eruptions or mass-transfer to a companion star may be needed. The fact that no WC stars and only two WO stars are known at metallicities below that of the LMC suggests that their formation at low metallicities requires a very specific evolutionary history. However, if these stars do form, we find that their winds are relatively strong for their metallicity.

\subsection{Hydrogen-free WN stars}

To assess in which domain of parameter space the new mass-loss prescription remains valid, we evaluate the predicted mass-loss rates of hydrogen-free WN stars (with $X=0$ and $Y \ga 0.98$) at various metallicities. 

\begin{figure*}
\epsscale{0.7}
\plotone{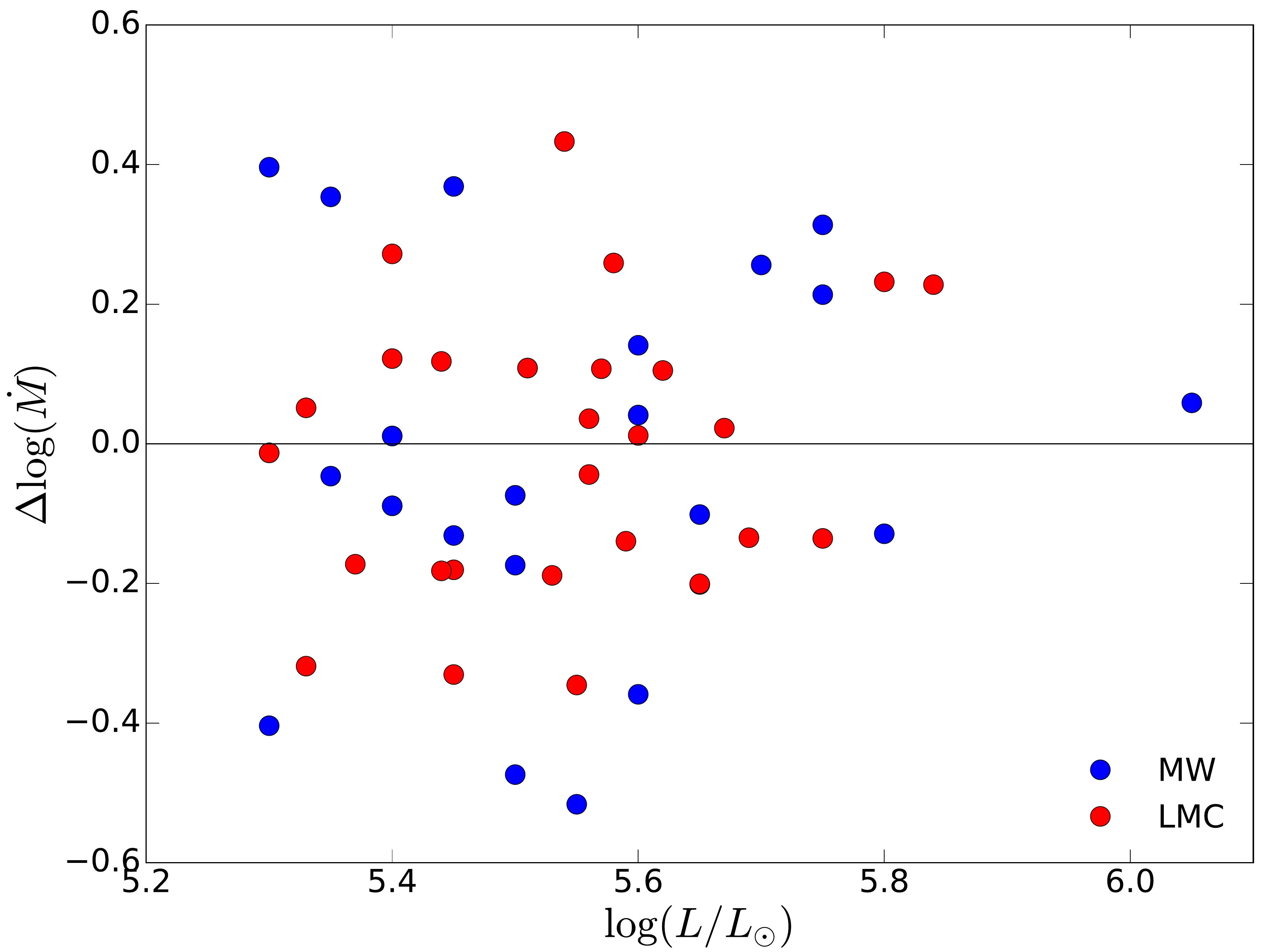}
\caption{Residuals of the mass-loss rates ($\Delta\log{\dot{M}} = \log{\dot{M}_{\mathrm{observed}}} - \log{\dot{M}}_{\mathrm{predicted}}$) computed for hydrogen-free WN stars, taking into account the stronger metallicity dependence of Equation~\ref{eq:recipeWN}. Stellar parameters from \citet[][MW]{hamann2006} and \citet[][LMC]{hainich2014}. \label{fig:mdot_WN}}
\end{figure*}

For the hydrogen-free, presumed single Galactic WN stars analysed by \cite{hamann2006}, our prescription provides mass-loss rates with an accuracy comparable to the NL00 prescription, i.e. with a standard deviation of 0.2 dex (see Figure~\ref{fig:mdot_WN}). For the presumed-single hydrogen-free WN stars in the Large Magellanic Cloud studied by \cite{hainich2014}, the mass-loss rates are significantly over-predicted, suggesting that the dependence on initial metallicity of Equation~\ref{eq:recipe} does not hold for WN stars. This is in line with the results from \cite{hainich2015}, who derive an empirical metallicity dependence for all WN stars (including those with hydrogen) using Galactic, M31, LMC, and Small Magellanic Cloud (SMC) WN stars. They find $\dot{M} \propto Z_{\mathrm{Fe}}^{1.4}$, thus a much stronger dependence on initial metallicity. 

To assess if Equation~\ref{eq:recipe} would be valid with a different dependence on initial metallicity, we fit Equation~\ref{eq:mdot} to the results for hydrogen-free WN stars in the Milky Way and LMC from \cite{hamann2006} and \cite{hainich2014}, keeping $A$, $B$, and $C$ fixed to the values of Equation~\ref{eq:recipe}. This approach is motived by the good results for the Galactic hydrogen-free WN stars, where the dependence on initial metallicity drops out. This indicates that the luminosity dependence holds, as, for hydrogen-free WN stars, $Y$ does not vary between stars with the same initial metallicity \citep[e.g., ][]{hamann2006}. We find a metallicity dependence of $\dot{M} \propto Z_{\mathrm{Fe}}^{1.3\pm0.2}$, in excellent agreement with the results from \cite{hainich2015}. Thus, the mass-loss rates of hydrogen-free WN stars can be described as

\begin{align}
\log{\dot{M}} = -9.20(\pm0.35) + 
0.85(\pm0.06)\log{\frac{L}{L_{\odot}}} \nonumber\\
+ 0.44(\pm0.08)\log{Y} + 
1.3(\pm0.2)\log{\frac{Z_{\mathrm{Fe}}}{Z_{\mathrm{Fe}, \odot}}}\label{eq:recipeWN}.
\end{align}

\noindent With this dependence on metallicity, the prescription reproduces the mass-loss rates of the hydrogen-free LMC WN stars with a standard deviation of 0.2 dex and no systematic offset, i.e. with an accuracy comparable to that of the Galactic case (see Figure~\ref{fig:mdot_WN}). 

A change in metallicity dependence between WN and WC stars is prediced by \cite{vink2005}, who find that in the metallicity range discussed here, the exponent of the $Z_{\mathrm{Fe}}$ dependence decreases from 0.86 for hydrogen-poor WN stars to 0.66 for WC stars. However, while the trend is in the right direction, the amplitude of the change in the exponent that reproduces the observations is about a factor of five larger than theory predicts. Combined with the shallow dependence on $Y$, this likely indicates that as the carbon abundance increases, this element becomes a more important driver of the wind.

We also compared our findings to WN stars that do contain hydrogen in the Galaxy, LMC, and SMC. For the sample of seven SMC WN stars studied by \cite{hainich2015}, spanning $X = 0.2$-$0.5$, our prescription matches the observed mass-loss rates with an accuracy comparable to that of the hydrogen-free WN stars in the Galaxy and LMC. However, for the WN stars with hydrogen in the Galaxy and LMC the results are poorer, the more so for sources with higher $X$.

We conclude that our mass-loss prescription behaves well for hydrogen-free WN stars, provided that the steeper metallicity dependence of $\dot{M} \propto Z_{\mathrm{Fe}}^{1.3}$ is taken into account. For WN stars that have a significant surface hydrogen mass fraction our prescription is less accurate.

\section{Summary}\label{sec:summary}

We have presented a new prescription for the mass-loss rates of hydrogen-free Wolf-Rayet stars as a function of their luminosity, surface abundances, and metallicity. The prescription (Equations~\ref{eq:recipe} and \ref{eq:recipeWN}) is based on the derived mass-loss rates of WC and WO stars in the Milky Way, Large Magellanic Cloud, and IC1613 galaxies. Equation~\ref{eq:recipe} is valid for hydrogen-free Wolf-Rayet stars with surface helium mass fractions $Y \la 0.98$ with a precision of $\sigma = 0.06$ dex. Equation~\ref{eq:recipeWN} is valid for hydrogen-free Wolf-Rayet stars with helium mass fractions $Y \ga 0.98$, albeit with larger residuals ($\sigma = 0.2$ dex, comparable to the NL00 prescriptions). In practice this means that the recipe is valid for all WC and WO stars, and for the hydrogen-free ($X=0$) subset of WN stars. Future implementation in evolutionary codes will allow to quantify the impact on the duration of late evolutionary stages, and the nature and properties of the final supernova explosion and compact remnant left behind.

\acknowledgments
We are grateful to dr. Ehsan Moravveji for useful discussions.

\bibliographystyle{apj}


\end{document}